\documentstyle[12pt]{article}

\begin{document}

\begin{titlepage}

\title{Correspondence between the XXZ model in roots 
       of unity and the one-dimensional quantum Ising chain 
       with different boundary conditions.}
\author{{\sc F.C.~Alcaraz}\\
{\em Universidade Federal de S\~ao Carlos}\\
{\em 13565-905,S\~ao Carlos,SP Brazil}
\vspace{0.5cm}\\
       {\sc A.A.~Belavin} and 
       {\sc R.A.~Usmanov}\\
{\em Landau Institute for Theoretical Physics}\\
{\em Chernogolovka, Moscow region 142432, Russia}
\vspace{0.5cm}}

\end{titlepage}
       
\maketitle

\begin{abstract}
We consider the integrable XXZ model with special open boundary 
conditions that renders its Hamiltonian ${SU(2)}_q$ symmetric, and 
the one-dimensional quantum Ising model with four different boundary 
conditions. We show that for each boundary condition the Ising 
quantum chain is exactly given by the Minimal Model of integrable 
lattice theory $LM(3, \, 4)$. This last theory is obtained as the 
result of the quantum group reduction of the XXZ model at anisotropy 
$\Delta=(q + q^{-1})/2=\sqrt{2}/2$, with a number of sites in 
the latter defined by the type of boundary conditions.
\end{abstract}

\section{Introduction}

Finite-width transfer matrices for four different boundary 
conditions on the cylinder were defined and studied in 
\cite{kn:gnus1}, \cite{kn:gnus2}, where it was shown that 
in each specific case, these matrices, which depend on a spectral 
parameter, form a commuting family and satisfy the same 
functional equation, which allows the evaluation of their eigenvalues.

As shown in \cite{kn:gnus14}, the logarithmic derivative of 
the above transfer matrices computed at the zero value of the spectral 
parameter is exactly the Hamiltonian of the one-dimensional quantum 
Ising chain with some boundary terms. The latter are different 
for each type of boundary conditions of the two-dimensional model. 
Thus there is a direct connection between the critical Ising model 
on a two-dimensional lattice and the one-dimensional quantum Ising 
chain.

Based on numerical analysis on finite chains in \cite{kn:gnus3}, 
\cite{kn:gnus9}, it was realized the remarkable fact that some 
of the eigenenergies of the XXZ Hamiltonian with the boundary 
condition that renders this Hamiltonian $SU(2)_q$ symmetric 
at the anisotropy $\Delta = (q+q^{-1})/2=\sqrt{2}/2$ exactly coincide 
with some of the eigenenergies of the quantum Ising chain.

The notion of quantum group reduction of the integrable XXZ 
model with open boundary conditions (OBC) in roots of unity 
was introduced in \cite{kn:gnus4}. The model given as a result 
of Quantum Group reduction is denoted by $LM(p, \, p+1)$ ($p$ 
is a chain parameter) and called the Minimal Model of integrable 
lattice theory \cite{kn:gnus6}. The thermodynamic limit 
($N \rightarrow \infty$) of $LM(p, \, p+1)$ is $M(p, \, p+1)$, 
the ordinary Minimal Model of CFT with the Virasoro central charge 
$c=1-{6}/{p(p+1)}$.

The integrability of the XXZ model with OBC manifests 
itself in the presence of a commuting family of transfer matrices 
found by Sklyanin \cite{kn:gnus11}, \cite{kn:gnus12}. 
Sklyanin's transfer matrices of the XXZ model after Quantum 
Group reduction also satisfy some functional equations 
as shown in \cite{kn:gnus6}. In the case $p=3$ these equations 
coincide with the functional equations for the Ising transfer 
matrices. This fact supports an equivalence of $LM(3, \, 4)$ 
and the Ising model.

In the present paper, we show that, in fact, these
two models exactly coincide for all types of boundary conditions 
on the Ising lattice introduced in \cite{kn:gnus2}. Namely, 
the following statements are true:

\begin{enumerate}

\item[1.] The XXZ chain with an odd number of sites ($2L+1$) after the 
quantum group reduction of the configuration space is equivalent 
to the $L$-site Ising chain with mixed boundary conditions.

\item[2.] The configuration space of the $2L$-site XXZ chain after the 
quantum group reduction can be decomposed into a direct sum of two 
subspaces with the same dimension $2^{L-1}$. These subspaces form 
two different representations of the Temperley--Lieb algebra 
${\cal T}_{2L-1}$ 
and are eigensubspaces of the Casimir operator of $U_q(sl(2))$. 
We'll denote $V_0$ and $V_1$  the subspaces  
corresponding to the eigenvalues ${(S^2)}_q =\sqrt{2}$ and 
${(S^2)}_q =0$, respectively. \
Then the identification must be done as follows:
\begin{enumerate}
\item[a.] the whole complex of eigenvalues of the Hamiltonian of the 
$2L$-site XXZ chain computed on vectors from $V_0$ coincide with 
the spectrum of the $(L-1)$-site Ising chain with the boundary 
conditions $(++)$ if $L$ is even and the boundary conditions $(+-)$ 
if $L$ is odd;

\item[b.] the whole complex of eigenvalues of the Hamiltonian of the 
$2L$-site XXZ chain computed on vectors from $V_1$ coincide with 
the spectrum of the $(L-1)$-site Ising chain with the boundary 
conditions $(++)$ if $L$ is odd and the boundary conditions $(+-)$ 
if $L$ is even.
\end{enumerate}

\item[3.] The spectrum of the $L$-site Ising chain with free 
boundary conditions coincide with the united spectrum of the 
$(L-1)$-sites Ising chains with the boundary conditions $(++)$ 
and $(+-)$.
\end{enumerate}

The plan of the paper is as follows. In part 2 we define the 
Ising model with different cylindrical boundary conditions. In part 
3, we introduce families of transfer matrices for four types of 
boundary conditions according to \cite{kn:gnus2} and find the 
connection between these transfer matrices and Hamiltonians of the 
one-dimensional Ising chains. In part 4, we consider the integrable XXZ
model with special open boundary conditions. In part 5, we then 
investigate two realizations of the Temperley--Lieb algebra in terms 
of dynamic variables of the Ising and XXZ chains. In part 6, we 
identify the Ising chain and $LM(3, \, 4)$. In part 7, we formulate 
some questions for future. In appendices A and B we show some useful 
relations used in part 5.

\section{The basic definitions}

We consider two types of two-dimensional Ising lattices defined as 
follows \cite{kn:gnus2}:
\begin{enumerate}
\item[1.] a finite-width square lattice rotated by 45 degrees with each
row having $L$ or $L-1$ faces;

\item[2.] a similar lattice but each row has $L$ faces.
\end{enumerate}

Vertically, both lattices have columns of $L'$ faces. We 
identify the first row of faces with the $(L'$$+$$1)$th row (the 
cylindrical boundary conditions). A lattice of the first type 
is denoted by the symbol ${\cal L}$ and a lattice of the second 
type by the symbol $\cal L'$ (see the figure). Lattice $\cal L$ 
consists of $2L$ and $\cal L'$ of $2L$+$1$ zigzagging columns. 
Let us denote this number by $N$; i. e., $N=2L$ for ${\cal L}$ 
and $N=2L+1$ for ${\cal L'}$. Further we take $L$ to be fixed.

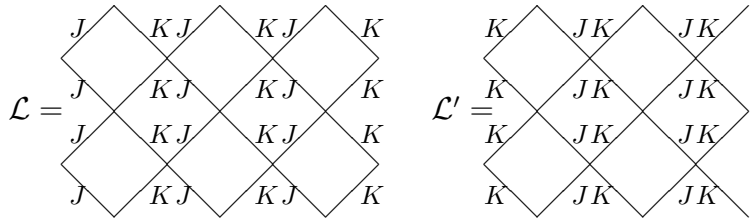
\begin{center}
\begin{picture}(300,80)(0,0)

\put(0,37){$\cal L=$}
\put(20,60){\line(1,-1){60}}
\put(40,80){\line(1,-1){80}}
\put(80,80){\line(1,-1){60}}
\put(120,80){\line(1,-1){20}}
\put(20,20){\line(1,-1){20}}
\put(20,60){\line(1,1){20}}
\put(20,20){\line(1,1){60}}
\put(40,0){\line(1,1){80}}
\put(80,0){\line(1,1){60}}
\put(120,0){\line(1,1){20}}
{
  \footnotesize
  \put(23,5){$J$}
  \put(23,28){$J$}
  \put(23,45){$J$}
  \put(23,68){$J$}
  \put(53,5){$K$}
  \put(53,28){$K$}
  \put(53,45){$K$}
  \put(53,68){$K$}
  \put(63,5){$J$}
  \put(63,28){$J$}
  \put(63,45){$J$}
  \put(63,68){$J$}
  \put(93,5){$K$}
  \put(93,28){$K$}
  \put(93,45){$K$}
  \put(93,68){$K$} 
  \put(103,5){$J$}
  \put(103,28){$J$}
  \put(103,45){$J$}
  \put(103,68){$J$}
  \put(133,5){$K$}
  \put(133,28){$K$}
  \put(133,45){$K$}
  \put(133,68){$K$}
}

\put(160,37){${\cal L}'=$}
\put(180,60){\line(1,-1){60}}
\put(200,80){\line(1,-1){80}}
\put(240,80){\line(1,-1){40}}
\put(180,20){\line(1,-1){20}}
\put(180,60){\line(1,1){20}}
\put(180,20){\line(1,1){60}}
\put(200,0){\line(1,1){80}}
\put(240,0){\line(1,1){40}}

{
  \footnotesize
  \put(180,5){$K$}
  \put(180,28){$K$}
  \put(180,45){$K$}
  \put(180,68){$K$}
  \put(213,5){$J$}
  \put(213,28){$J$}
  \put(213,45){$J$}
  \put(213,68){$J$}
  \put(220,5){$K$}
  \put(220,28){$K$}
  \put(220,45){$K$}
  \put(220,68){$K$}
  \put(253,5){$J$}
  \put(253,28){$J$}
  \put(253,45){$J$}
  \put(253,68){$J$}
  \put(260,5){$K$}
  \put(260,28){$K$}
  \put(260,45){$K$}
  \put(260,68){$K$}
}

\end{picture}
\end{center}

On these lattices we can define the Ising model by attaching at each 
lattice site a spin variable taking the values +$1$ or $-1$. Because, 
in contrast to the toroidal model, the boundary spins in some 
row are not  equal, we can consider different
boundary conditions:

$++$: we choose the lattice $\cal L$ and fix the spins at the
left and right boundaries to be +1;

$+-$:  we choose the lattice $\cal L$ and fix the left 
boundary spins to be +1 and the right boundary spins to be $-1$;

{\it mixed boundaries}: we choose the lattice $\cal L'$ 
and fix the left boundary spins to be +1, but place no 
restriction on the right boundary spins;

{\it free boundaries}: we choose the lattice $\cal L$ with no 
restrictions on the boundary spins.

The lattices with the other possible boundary conditions, 
like, for example, $--$ or $-+$, are clearly related with 
the above ones.

\section{Transfer matrices}

The set of spins in some row that are not fixed by boundary 
conditions are denoted by $\Phi$. We define transfer matrices 
as follows \cite{kn:gnus2}:\\ 
1. {\it for free boundaries},
$$
T_{\Phi,{\Phi}'}=\sum_{\Phi''}\exp\left(J\sum_{j=1}^L{\sigma}{''_j}
({\sigma}_j+{\sigma}{'_j})+K\sum_{j=1}^L{\sigma}{''_{j+1}}
({\sigma}_j+{\sigma}{'_j})\right);
$$
2. {\it for fixed boundary conditions},
$$
T_{\Phi,{\Phi}'}=\sum_{\Phi''}\exp\left(
2J{\sigma}{''_1}
+K\sum_{j=1}^{L-1}{\sigma}{''_j}
({\sigma}_j+{\sigma}{'_j})+J\sum_{j=1}^{L-1}{\sigma}{''_{j+1}}
({\sigma}_j+{\sigma}{'_j})+2{\tau}K{\sigma}{''_{L}}\right)
$$
where $\tau=1$ or $\tau = -1$, for the boundary conditions 
$(++)$ or $(+-)$ respectively;\\
3. {\it for mixed boundary conditions}:
$$
T_{\Phi,{\Phi}'}=\sum_{\Phi''}\exp\left(2K{\sigma}{''_1}+
J\sum_{j=1}^L{\sigma}{''_j}
({\sigma}_j+{\sigma}{'_j})+K\sum_{j=1}^L{\sigma}{''_{j+1}}
({\sigma}_j+{\sigma}{'_j})\right).
$$
Here, $\Phi$, $\Phi''$, and $\Phi'$ are three successive sets of spins,
and we sum over $\Phi''$. 
Transfer matrices are defined on the lattice $\cal L$ in the 
case of free and fixed boundary conditions and on the lattice 
$\cal L'$ in the case of mixed boundary conditions.

The transfer matrices defined above have a 
remarkable property: 
$$
T(J,K)T(J',K')=T(J',K')T(J,K)
$$
(that is they commute with each other) if
\begin{equation}
\sinh(2K)\sinh(2J)=\sinh(2K')\sinh(2J').
\label{eq:Sh*Sh}
\end{equation}

We consider the critical Ising model; therefore 
$\sinh(2J)\sinh(2K)=1$. Following Baxter, we use 
the parameterization:
$$
\sinh(2J)=\cot(u), \qquad   \sinh(2K)=\tan(u),
$$
where $0<u<\pi/2$ is a spectral parameter.

As shown in \cite{kn:gnus2}, the transfer matrices $T(u)$ 
satisfy the following functional equation
\begin{equation}
T(2u)T(2u+\frac{\pi}{2})=\frac{{\cos}^{2(N+1)}(2u)-{\sin}^{2(N+1)}(2u)}{\cos(4u)}
\frac{2^{N}{(-1)}^{L}}{({{\sin}(2u){\cos}(2u))}^{N}}
\label{ur:IS}
\end{equation}
in the case of fixed boundary conditions and
\begin{equation}
T(2u)T(2u+\frac{\pi}{2})=\frac{{\cos}^{2(N+1)}(2u)-{\sin}^{2(N+1)}(2u)}{\cos(4u)}
\frac{2^{N+2}{(-1)}^{L}}{({{\sin}(2u){\cos}(2u))}^{N}}
\label{ur:IS_1}
\end{equation}
in the case of free and mixed boundary conditions, 
where $N$ and $L$ are introduced in part 2.

We now consider the Hamiltonian of the one-dimensional quantum Ising 
chain with $L$ sites,
\begin{equation}
H_{\rm Ising}^F(L)=\sum_{i=1}^{L-1}{\sigma_j^z}{\sigma_{j+1}^z}+
\sum_{i=1}^{L}{\sigma_j^x}.
\end{equation}

This Hamiltonian is related to the transfer matrix $T(u)$ of 
the Ising model with free boundary conditions by
\begin{center}
\begin{picture}(200,20)
\put(0,5){$H_{\rm Ising}^F(L)=T^{-1}(0)\dot{T}(0) - $ singular terms,}
\end{picture}
\end{center}
where $T^{-1}(u)$ is an inverse matrix and $\dot{T}(u)$ is the 
derivative of the matrix $T(u)$ with respect to its parameter.

For other boundary conditions the logarithmic derivative in zero is
slightly modified \cite{kn:gnus14}:\\
1. for $(++)$, we obtain the Hamiltonian of the $(L$$-$$1)$-site Ising
chain
$$
H_{\rm Ising}^{++}(L)=H_{\rm Ising}^F(L-1)+{\sigma}_1^z+{\sigma}_{L-1}^z,
$$
2. for $(+-)$, we obtain the Hamiltonian of the $(L$$-$$1)$-site Ising
chain
$$
H_{\rm Ising}^{+-}(L)=H_{\rm Ising}^F(L-1)+{\sigma}_1^z-{\sigma}_{L-1}^z,
$$
3. for mixed boundary conditions, we obtain the Hamiltonian of 
the $L$-site Ising chain
$$
H_{\rm Ising}^{M}(L)=H_{\rm Ising}^F(L)+{\sigma}_1^z,
$$
where $H_{\rm Ising}^F$ is the Hamiltonian of the model with free 
boundary conditions. 
The proof of this can be found in \cite{kn:gnus14}.

\section{Minimal Models of Integrable Lattice Theory}

We consider the one-dimensional XXZ chain with free boundary conditions
\cite{kn:gnus3}
\begin{equation}
\begin{array}{lll}
\displaystyle
H_{XXZ}&=&\displaystyle \sum_{n=1}^{N-1}[{\sigma}_{n}^{+}{\sigma}_{n+1}^{-}+
{\sigma}_{n}^{-}{\sigma}_{n+1}^{+}+\frac{q+q^{-1}}{4}
{\sigma}_{n}^{z}{\sigma}_{n+1}^{z}\\
&&\displaystyle
\;\;\;\;\;\;+\frac{q-q^{-1}}{4}({\sigma}_{n}^{z}-{\sigma}_{n+1}^{z})],
\\&&\\
{\sigma}_{n}^{\pm}&=&1 \otimes ... \otimes {\sigma}^{\pm}
\otimes ... \otimes 1,\\ \displaystyle
{\sigma}_{n}^{z}&=&1 \otimes ... \otimes {\sigma}^{z} \otimes ...
\otimes 1,\\ \displaystyle
{\sigma}^{+}&=&

\left(
\begin{array}{cc}
0&1 \\ 0&0
\end{array}
\right), \;\;\;\;\;

{\sigma}^{-}= 
\left(
\begin{array}{cc}
0&0 \\ 1&0
\end{array}
\right), \;\;\;\;\;

{\sigma}^{z}=

\left(
\begin{array}{cc}
1&0 \\ 0&-1
\end{array}
\right).

\end{array}
\end{equation}
In the thermodynamic limit ($N \rightarrow \infty$) this 
Hamiltonian is gapless for 
$-1 \leq \Delta = (q+q^{-1})/2 \leq 1$.  
The model has remarkable properties. 
Besides the $U(1)$ symmetry, translated into its commutation 
with the $z$-component of the total magnetization 
$\displaystyle S^z=\frac12 \sum_{i=1}^N {\sigma}_i^z$, it 
was also shown in \cite{kn:gnus4} that the Hamiltonian $H_{XXZ}$ 
commutes with the quantum group $U_q({\it sl}(2))$ including its 
generators X, Y and H, which are defined as
$$
\begin{array}{lll}
X&=&\displaystyle
\sum_{n=1}^{N}q^{\frac12({\sigma}_{1}^{z}+...+{\sigma}_{n-1}^{z})}
{\sigma}_{n}^{+}q^{-\frac12({\sigma}_{n+1}^{z}+...+{\sigma}_{N}^{z})},\\
Y&=& \displaystyle
\sum_{n=1}^{N}q^{\frac12({\sigma}_{1}^{z}+...+{\sigma}_{n-1}^{z})}
{\sigma}_{n}^{-}q^{-\frac12({\sigma}_{n+1}^{z}+...+{\sigma}_{N}^{z})},\\
H&=&\displaystyle
\sum_{n=1}^{N}\frac{{\sigma}^z_n}{2},
\end{array}
$$
and satisfies the relations
\begin{equation}
[H,X]=X, \;\;\;\; [H,Y]=-Y, \;\;\;\;
[X,Y]=\frac{q^{2H}-q^{-2H}}{q-q^{-1}}.
\end{equation}
Furthermore, the densities
$$
H_{n}=\displaystyle {\sigma}_{n}^{+}{\sigma}_{n+1}^{-}+
{\sigma}_{n}^{-}{\sigma}_{n+1}^{+}+\frac{q+q^{-1}}{4}
{\sigma}_{n}^{z}{\sigma}_{n+1}^{z}
+\frac{q-q^{-1}}{4}({\sigma}_{n}^{z}-{\sigma}_{n+1}^{z}),
$$
where $n=1,...,N-1$, also commute with the quantum group 
$U_q({\it sl}(2))$.

Because of the quantum group symmetry the spectrum of the
Hamiltonian can be classified according to the representation 
theory of the algebra $U_q({\it sl}(2))$.

The representation theory of $U_q({\it sl}(2))$ has been studied in detail
in the case where $q$ is not a root of unity. In this case, its 
representations are equivalent to those of the usual $U({\it
sl}(2))$, that is, the configuration space ${({\bf C}^2)}^N$ of the 
spin chain can be split into a direct sum of irreducible highest-weight 
representations ${\rho}_j$ ($j$ is the highest weight), which are in 
one-to-one correspondence with the ordinary ${\it sl}(2)$ 
representations. For example, in case $N=4$, where ${({\bf C}^2)}^4$ 
can be decomposed into ${\rho}_2+3{\rho}_1+2{\rho}_0$.

We study the case where $q^{p+1}=-1$ \cite{kn:gnus4}, \cite{kn:gnus7}. 
In this case, the generators X and Y are nilpotent:
\begin{equation}
X^{p+1}=0, \qquad Y^{p+1}=0.
\end{equation}
We consequently obtain a very different picture of 
the representations.

For example if $q^{4}=-1$ and $N=4$ then $X^4=0$ and $Y^4=0$. 
The space ${({\bf C}^2)}^4$ now decomposes into the sum of one "bad" 
eight-dimensional representation $({\rho}_2,{\rho}_1)$ of type $I$ and 
four other "good" representations $2{\rho}_1+2{\rho}_0$ of type $II$ 
\cite{kn:gnus4}. There is an isomorphism of the type-$II$ representations 
and the ordinary $U({\it sl}(2))$ ones. The type-$I$ representation can be
considered as the result of gluing two representations ${\rho}_2$ and
${\rho}_1$. It is indecomposable but is not irreducible (it contains 
three-dimensional invariant subspace).

In the general case with $q^{p+1}=-1$ \cite{kn:gnus4}, \cite{kn:gnus7} 
the configuration space splits into the sum of "bad" type-$I$ 
representations with the highest weights $S_z \geq p/2$ and "good" type 
$II$ representations with highest weights $S_z<p/2$ which are 
simultaneously not subspaces of some "bad" ones. The highest-weight 
vectors $v_j$ of the good representations can be characterized by the 
condition
\begin{equation}
v_j \in V_p \equiv Ker X/Im X^p.
\end{equation}
Because $U_q({\it sl}(2))$ commutes with $H_{XXZ}$, we can 
normally restrict its action on the space $V_p$. The model 
resulting from this quantum group reduction model is called 
$LM(p, \; p+1)$ \cite{kn:gnus6}.

The number of representations with the highest weight $j<\frac{p}{2}$ 
in the decomposition of ${({\bf C}^2)}^N$ equals the number of restricted 
paths of length $N$ beginning at zero and ending at $j$. 
The restriction means that a path cannot cross the straight lines 
$j=0$ and $j=p/2$. For example, if $p=3$ and $N$ is odd, 
then only the paths ending at $j=1/2$ are permissible, and 
the number of paths is therefore $2^{N-1}$.

A so-called Sklyanin transfer matrix \cite{kn:gnus11}, \cite{kn:gnus12}:
$$
T_{1/2}(u)={(-1)}^N tr(e^{-{\sigma}^z(u+\eta)}L(u)e^{{\sigma}^z u}L^{t \otimes
t}(u)).
$$
is related to Hamiltonians of the XXZ chain. 
Here $L(u)$ is a monodromy matrix
$$
L(u)=R_N(u)...R_1(u).
$$
and $R(u)$ is given by the expression
$$
\left ( 
\begin{array}{cc}
\left(
\begin{array}{cc}
\sin(u+\eta) & 0 \\
0 &  \sin(u)
\end{array}
\right) & 

\left(
\begin{array}{cc}
0 & 0 \\
 \sin(\eta) & 0
\end{array}
\right) \\ 

\vspace{-2mm}\\

\left(
\begin{array}{cc}
0 &  \sin(\eta) \\
0 & 0
\end{array}
\right) & 

\left(
\begin{array}{cc}
 \sin(u) & 0 \\
0 &  \sin(u+\eta)
\end{array}
\right) 
\end{array}
\right )
$$
The relation
$$
H_{XXZ}=\left.\frac{\sin(\eta)}{2}\frac{d \log
T_{1/2}(u)}{du}\right |_{u=0}
+\frac{{\sin}^2(\eta)}{2\cos
(\eta)}-\frac{N}{2}\cos(\eta)
$$
holds. The transfer matrices commute with each other under different 
values of the spectral parameter and therefore with the Hamiltonian:
$$
[T_{1/2}(u),T_{1/2}(v)]=0, \qquad [T_{1/2}(u),H_{XXZ}]=0.
$$
As shown in \cite{kn:gnus6}, after the quantum group reduction, 
$T_{1/2}(u)$ satisfies the functional equation
\begin{equation}
T_{1/2}(u)T_{1/2}(u+\frac{\pi}{4})=2^{-2N+1} \frac{{\cos}^{2(N+1)}(2u)-
{\sin}^{2(N+1)}(2u)}{\cos(4u)}.
\label{ur:Skl}
\end{equation}
Comparing the equations (\ref{ur:IS}) and (\ref{ur:Skl}), 
we can see that in the case of even $N$ and fixed boundary 
conditions, the matrices $T_{1/2}(u)$ and  \\
$2^{1/2-2N}{(\sin(4u))}^L T_{\rm Ising}(2u)$ 
satisfy the same functional equation. (Here, the transfer 
matrix of the Ising model with fixed boundary conditions 
is denoted by $T_{\rm Ising}(u)$ rather than $T(u)$ as it was 
in (\ref{ur:IS}).) Their eigenvalues hence coincide, and 
the two matrices are therefore equivalent.

In the case of even $N$ and free boundary conditions, the 
matrices $T_{1/2}(u)$ and 
$2^{-1/2-2N}{(\sin(4u))}^LT_{\rm Ising}(2u)$ also 
satisfy the same functional equation. Here, 
$T_{\rm Ising}(u)$ is the transfer matrix of the Ising model 
with free boundary equations.

Similarly, in the case of odd $N$, that is, for mixed boundary 
conditions, the matrices $T_{1/2}(u)$ and 
$2^{1/2-2N}{(\sin(2u))}^{L+1}{(\cos(2u))}^L T_{\rm Ising}(2u)$ 
satisfy the same functional equation. Here, $T_{\rm Ising}(u)$ is 
the transfer matrix of the Ising model with mixed boundary 
conditions.

Once again, we emphasize that while in the Ising model 
we have $L$ spins in the related XXZ chain we have $N=2L$ 
spins if the boundary conditions in the Ising model are free 
or fixed, and $N=2L+1$ in the case where the Ising chain 
has mixed boundary conditions. This means that the dimension 
of $T_{1/2}$, which is $2^N$ is always bigger than the 
corresponding dimension $2^L$ of the related $T_{\rm Ising}$.

That the transfer matrices in the two models satisfy 
functional equations of similar form suggests the 
essential identity of the models. In what follows, 
we establish this identity of the XXZ and Ising models 
and show how it can be realized.

\section{Temperley-Lieb algebra}

We recall the definition of the Temperley--Lieb algebra as an 
algebra with the generators $e_i,\; i=1,...,n,$ satisfing 
the relations:
\begin{equation}
e_i e_{i \pm 1} e_i = e_i, \;\;\;\;\;\;\; e_i^2=(q+q^{-1})e_i,
\;\;\;\;\;\;\; [e_i,e_j]=0, \qquad |i-j|>1.
\label{eq:Def_TL}
\end{equation}
The Temperley--Lieb algebra with $n$ generators is 
denoted by ${\cal T}_n$. Further, we assume that $q=e^{i \pi /4}$.

There are realizations of the Temperley--Lieb algebra in terms of the 
dynamic variables of Ising and XXZ model. Namely, it can be verified that

\begin{enumerate}
\item[1.] The expressions
$$
\begin{array}{llll}
\displaystyle
e_{2i-1}&=&\displaystyle
\frac{1}{\sqrt{2}}({\sigma}_{i}^{z}{\sigma}_{i+1}^{z}+1),& 
i=1,...,L\\ 
\vspace{-2mm}\\
\displaystyle
e_{2i}&=&\displaystyle \frac{1}{\sqrt{2}}({\sigma}_{i}^x+1),& 
i=1,...,L-1,
\end{array}
$$
give a realization of ${\cal T}_{2L-1}$;

\item[2.] the expressions
$$
\begin{array}{llll}
\displaystyle
e_1&=&\displaystyle \frac{1}{\sqrt{2}}({\sigma}_{1}^z+1),\\
\vspace{-2mm}\\
\displaystyle
e_{2i}&=&\displaystyle
\frac{1}{\sqrt{2}}({\sigma}_{i}^{z}{\sigma}_{i+1}^{z}+1),&
i=1,...,L\\ 
\vspace{-2mm}\\
\displaystyle
e_{2i-1}&=&\displaystyle \frac{1}{\sqrt{2}}({\sigma}_{i}^x+1),&
i=2,...,L
\end{array}
$$
give a realization of ${\cal T}_{2L}$;

\item[3.] the expressions
$$
\begin{array}{llll}
\displaystyle
e_1&=&\displaystyle \frac{1}{\sqrt{2}}({\sigma}_{1}^z+1),\\
\vspace{-2mm}\\
\displaystyle
e_{2i-1}&=&\displaystyle
\frac{1}{\sqrt{2}}({\sigma}_{i}^{z}{\sigma}_{i+1}^{z}+1),& i=2,...,L,\\
\vspace{-2mm}\\
\displaystyle e_{2i}&=&\displaystyle 
\frac{1}{\sqrt{2}}({\sigma}_{i}^x+1), & i=1,...,L-1\\
\vspace{-2mm}\\  
\displaystyle
e_{2L-1}&=&\displaystyle \frac{1}{\sqrt{2}}({\sigma}_{L-1}^z+1). 
\end{array}
$$
give a realization of ${\cal T}_{2L-1}$;

\item[4.] the expressions
$$
\begin{array}{llll}
\displaystyle
e_1&=&\displaystyle \frac{1}{\sqrt{2}}({\sigma}_{1}^z+1),\\
\vspace{-2mm}\\
\displaystyle
e_{2i-1}&=&\displaystyle
\frac{1}{\sqrt{2}}({\sigma}_{i}^{z}{\sigma}_{i+1}^{z}+1),& i=2,...,L,\\
\vspace{-2mm}\\
\displaystyle e_{2i}&=&\displaystyle 
\frac{1}{\sqrt{2}}({\sigma}_{i}^x+1), & i=1,...,L-1\\  
\vspace{-2mm}\\
\displaystyle
e_{2L-1}&=&\displaystyle \frac{1}{\sqrt{2}}(-{\sigma}_{L-1}^z+1). 
\end{array}
$$
give a realization of ${\cal T}_{2L-1}$.
\end{enumerate}
Using these expressions, we see that Hamiltonian of the Ising model 
for all four boundary conditions is
$$
\displaystyle H_{\rm Ising}=\sum_{i=1}^{N-1}\left( e_i \sqrt{2} - 1
\right), 
$$
where $N=2L$ for free and fixed boundary conditions and 
$N=2L+1$ for mixed boundary conditions.

In terms of the dynamic variables of the XXZ model, the expressions
$$
e_i=-H_i + \frac{\sqrt{2}}{4},
$$
where $i=1,...,N-1$, give a realization of ${\cal T}_{N-1}$, and the 
Hamiltonain of the $N$-site XXZ chain is
$$
H_{XXZ}=\displaystyle -\sum_{i=1}^{N-1}\left(e_i - \frac{
\sqrt{2}}{4}\right).
$$

That the Hamiltonians of XXZ and Ising models have the same form in 
terms of generators of the Temperley--Lieb algebra supports the 
equivalence of the two models.

\section{Identification of $LM(3,\, 4)$ and the Ising \\ model}

\subsection{The XXZ chain with an odd number of sites and the Ising 
chain with mixed boundary conditions}

We consider the $(2L$$+$$1)$-site XXZ chain and the $L$-site Ising 
chain with mixed boundary conditions. The same 
algebra ${\cal T}_{2L}$ corresponds to each of them.

The densities 
$H_i$ of the Hamiltonian of the XXZ model commute 
with the quantum group $U_q({\it sl}(2))$. Therefore, the 
configuration space of the XXZ model after the quantum group reduction 
forms a representation of the algebra ${\cal T}_{2L}$ as it was before the 
reduction. This representation, whose vectors are in one-to-one 
correspondence to restricted paths (just such as in the 
RSOS model) of length $2L$$+$$1$ and height $2j=1$, 
has the dimension $2^L$ and is irreducible. 
The realization of the Temperley--Lieb algebra on the 
Ising configuration space gives the same representation. 
The equivalence of the two models is implied these facts 
and the fact that their Hamiltonians have the same form 
in terms of the generators of the Temperley--Lieb algebra.

This statement can be check 
numerically by comparing the eigenvalues of the operator
$$
\displaystyle \sum_{i=1}^{N-1} e_i
$$
computed on vectors from the configuration space of 
$LM(3, \, 4)$ and on vectors from the configuration space of 
the Ising model for small $L$.

\subsection{The XXZ chain with even number of sites and the Ising
chain with fixed boundary conditions}

We now consider the XXZ chain with $N=2L$ sites and the 
$(L$$-$$1)$-site Ising chain with fixed boundary conditions.
In the previous part, it was shown that the same algebra 
${\cal T}_{N-1}$ corresponds to each of them.

The dimension of the configuration space of each of the two 
Ising Hamiltonians in this case is half the dimension 
of the configuration space of $LM(3, \, 4)$ $V_3$. The 
space $V_3$ has the dimension $2^L$ and is decomposed into the sum 
of two subspaces. Each of them is an eigensubspace of the Casimir 
operator
$$
{(S^2)_q}=Y X + {\left( \frac{q^{H+1/2}-q^{-H-1/2}}{q-q^{-1}} \right)}^2-
{\left( \frac{q^{1/2}-q^{-1/2}}{q-q^{-1}} \right)}^2,
$$
one corresponding to the eigenvalue ${(S^2)}_q=0$, 
and the other corresponding to the eigenvalue ${(S^2)}_q=\sqrt{2}$. 
The first subspace is denoted by $V_0$ and the second 
by $V_1$. Because the Casimir operator commutes with the algebra 
${\cal T}_{2L-1}$, $V_0$ and $V_1$ form a representation of that algebra. 
The dimension of each subspace is $2^{L-1}$, which is 
exactly the dimension of the configuration space of each of 
the two Ising chains.

As shown in Appendix \ref{C},
$$
2^{2N} q^{-N} e^{-2iNu} T_{1/2}(u \rightarrow -i \infty)=
\left\{ {(q-q^{-1})}^2
{(S^2)}_q + (q+ q^{-1})\right\},
$$
and since the subpaces $V_0$ and $V_1$ are 
eigensubspaces of the matrix in the right side, 
we can hence see that $V_0$ and $V_1$ are 
simultaneously eigensubspaces of the matrix in the 
left side, with  eigenvalues $-1$ and $+1$, respectively. 

The densities $H_i$ of the Hamiltonian of the XXZ model commute
with the quantum group $U_q({\it sl}(2))$. Therefore, the 
configuration space of the XXZ model after the quantum group reduction 
forms a representation of the algebra ${\cal T}_{2L}$ as it was before the 
reduction. This representation, whose vectors are in one-to-one 
correspondence to restricted paths of length $2L$ and 
height $2j=0$ and $2j=2$, has the dimension $2^L$ and 
can be decomposed into the direct sum of two irreducible 
representations corresponding to two different heights of 
paths. The dimension of each representation equals $2^{L-1}$. 
The equivalence of the two models in each of the cases 
in question is implied these facts and the fact that their 
Hamiltonians have the same form in terms of the generators 
of the Temperley--Lieb algebra.

It was shown in \cite{kn:gnus2} that $T_{\rm Ising}(-i\infty)
=\pm 2^{L}$, where the upper sign corresponds to the boundary 
conditions $(++)$ and the lower sign corresponds to the 
boundary conditions $(+-)$. The matrices $2^{1-2N} {(\sin(4u))}^L 
T_{\rm Ising}(2u)$ and $T_{1/2}(u)$ become equivalent after the 
quantum group reduction. We can therefore write
$$
T_{1/2}(u)=2^{1-2N}{(\sin(4u))}^L T_{\rm Ising}(2u)
$$
in the $u\rightarrow -i\infty$ limit. Substituting 
$q=e^{i\pi /4}$, we obtain
\begin{equation}
\frac{1}{2^L}T_{\rm Ising}(-i\infty)={(-1)}^L \left\{ -\sqrt{2}{(S^2)}_q+1
\right\}.  
\label{eq:ravenstvo}
\end{equation}
Because the eigenvalues of the Casimir operator equal $0$ and 
$\sqrt{2}$, we can conclude that for even $L$, 
the configuration space of the Ising chain with the boundary 
conditions $(++)$ corresponds to $V_0$ and the configuration 
space of the Ising chain with the boundary conditons $(+-)$ 
corresponds to $V_1$. For odd $L$, the boundary conditions 
$(++)$ correspond to $V_1$ and the boundary conditions $(+-)$ 
correspond to $V_0$.

\subsection{The XXZ chain with an even number of sites and the 
Ising chain with free boundary conditions}

We now consider the XXZ chain with $N=2L$ sites and the $L$-site 
Ising chain with free boundary conditions. The algebra ${\cal T}_{2L-1}$ 
corresponds to each of them. The dimensions of the configuration 
spaces of the Ising chain and $LM(3, \, 4)$ both equal $2^L$.

In the case of free boundary conditions the Ising quantum 
chain has a Z(2) symmetry translated by the commutation of 
$H^F_{\rm Ising}$ with the parity operator C given by
$$
C={\sigma}^x_1{\sigma}^x_2...{\sigma}^x_L
$$
Hence, the configuration space of the Ising chain with free 
boundary conditions is decomposed into the sum of two sectors, 
corresponding to two eigenvalues of $C$. These two sectors are 
denoted by $C_{+}$ and $C_{-}$.

We show in appendix \ref{D} that $C$ is related to 
the limit of the transfer matrix of the $L$-site Ising 
model with free boundary conditions by
$$
C=\frac{1}{2^{L+1}}T_{\rm Ising}(-i \infty).
$$

As with fixed boundary conditions, we can 
obtain the identity
\begin{equation}
C=\frac{1}{2^{L+1}}T_{\rm Ising}(-i\infty)={(-1)}^{L} \left\{
-\sqrt{2}{(S^2)}_q+1
\right\},
\label{eq:ravenstvo1}
\end{equation}
from which we can see that there is a one-to-one 
correspondence between the sectors in the configuration 
space of the Ising chain and the subspaces $V_0$ and $V_1$. 
Namely, if $L$ is even, then the sector $C_{+}$ corresponds to 
$V_0$ and the sector $C_{-}$ to $V_1$. If $L$ is odd, then 
$C_{-}$ corresponds to $V_0$ and $C_{+}$ to $V_1$. 

\medskip
We have thus proved all statements formulated in the 
introduction.

\section{Some questions for future.}

We now formulate some questions for future, on which 
we'll try to answer in one of the following papers.

In the previous parts we have considered the critical 
Ising model and the homogeneous XXZ model and proved 
their equivalence. It seems our proof can be generalized 
on the case of the non-critical inhomogeneous Ising model 
given by the Hamiltonian
$$
H_{\rm Ising}(L)= \sum_{i=1}^{L-1}a_i {\sigma}^z_{i}
{\sigma}^z_{i+1} + \sum_{i=1}^L b_i {\sigma}^x_i,
$$
and the inhomogeneous XXZ model given by
$$
H_{XXZ}(N)=\sum_{i=1}^{N-1}c_i H_i,
$$
where $a_i$, $b_i$ and $c_i$ are arbitrary 
coefficients.

The inhomogeneous Hamiltonian XXZ commutes 
with the quantum \\ group  $U_q(sl(2))$, since 
the densities $H_i, \; i=1,...,N-1$ themselves 
commute with it.

Furthermore we can obtain the Hamiltonian XXZ 
by evaluating the logarithmic derivative of the 
Sklyanin's transfer matrix but now we must 
use its inhomogeneous version, that is, the 
monodromy matrix $L$ equals the product of 
$R$-matrices (as it was in part 4) evaluated 
at different values of the parameter $u$:
$$
L(u) = R_N(u-u_N)...R_2(u-u_2)R_1(u-u_1).
$$
It can be obtain that
$$
c_i = \frac{\sin \eta}{{\sin}^2 \eta - {\sin}^2  u_i},
$$
where $i=1,...,N$$-$$1$.

In particularly, in the case 
$$a_i = -1, \qquad b_i=-\lambda, \qquad 
\displaystyle c_i=\frac{\lambda + 1}{2} + 
(-1)^i \frac{\lambda - 1}{2},
$$
both Hamiltonians 
have the same form in terms of the generators of 
the Temperley--Lieb algebra:
$$
\begin{array}{l}
\displaystyle H_{\rm Ising}(L,\lambda)=-\sum_{i=1}^{N-1}
\left( \frac{\lambda + 1}{2} +(-1)^i \frac{\lambda - 1}{2} 
\right) \left( e_i \sqrt{2} - 1 \right),\\
\vspace{-2mm}\\
\displaystyle H_{XXZ}(N,\lambda) = - \sum_{i=1}^{N-1} \left(
\frac{\lambda + 1}{2} + (-1)^i \frac{\lambda - 1}{2}\right)
\left(e_i - \frac{\sqrt{2}}{4}\right),
\end{array}
$$
with the generators of the Temperley--Lieb algebra 
in terms of the dynamic variables of the Ising model 
given by the old expressions and in terms of 
the dynamic variables of the XXZ model given by
$$
e_i = - \frac{H_i}{(\lambda + 1)/2 + (-1)^i 
(\lambda - 1)/2}+\frac{q+q^{-1}}{4}, \qquad i=1,...,N-1.
$$
Consequently the Hamiltonians of the XXZ and the Ising 
models, properly normalized
\begin{equation}
{\tilde{H}}_{\rm Ising} (L, \lambda) = 
H_{\rm Ising}(L,\lambda)/{\sqrt{2}} - C_N \sqrt{2}/2
\label{eq:Ising_tilde}
\end{equation}
and
\begin{equation}
{\tilde{H}}_{XXZ} (N, \lambda) = H_{XXZ}(N, \lambda) - 
C_N \sqrt{2}/4,
\label{eq:XXZ_tilde}
\end{equation}
with
$$
C_N = \sum_{i=1}^N \left( \frac{\lambda + 1}{2} 
+(-1)^i \frac{\lambda - 1}{2} \right),
$$
are expressed in a similar form in terms of the 
generators of the Temperley--Lieb algebra. This 
fact supports an equivalence of the two models.

To illustrate we exhibit in table 1 the eigenenergies of 
the XXZ chain with $N$=$5$ sites and those of the $L$$=$$2$ 
Ising chain with mixed boundary condition, both models 
at $\lambda = 2$. The eigenenergies of the XXZ chain are 
separated according to their z-magnetization $S^z$, and 
the spin $S$ of the highest weights are also shown. The 
energy levels forming inhomogeneous $LM(3,4)$ have a 
superscript $(+)$, and we see their equality with the 
energy levels of the related Ising quantum chain. The 
level with an asterisk symbol, although having spin $S =1/2$ 
does not belong to inhomogenous $LM(3,4)$ since it is 
degenerated with another level with $S=5/2$.

In table 2 we illustrate relations between the two models 
by showing the eigenspectra of $\tilde H_{\rm Ising}^{++} (L=2, \lambda=2$), 
$\tilde H_{\rm Ising}^{+-} (L=2, \lambda=2$) and part of the eigenspectrum of 
$\tilde H_{XXZ}(N=6,\lambda=2)$. The eigenenergies of the XXZ chain 
are separated in the $S^z$ sectors and the spin $S$ of the highest weights 
are also shown. The energy levels forming $ V_0$ ($S=0$) and $ V_1$ 
($S = 1$) have a superscript $(0)$ and $(1)$, respectively. The 
eigenenergies having an asterisk does not belong to $ V_1$ since they 
are degenerated with the other eigenenergies with $S >1$. We see from 
this table that inhomogeneous $LM(3,\, 4)$ is obtained by gluing 
together the eigenspectra of the related Ising chains.

In table 3 we illustrate correspondences between the Ising 
model and inhomogeneous $LM(3, \, 4)$ by showing the eigenenergies of 
$\tilde H_{XXZ}(N=4,\lambda=2)$ and $\tilde H_{\rm Ising}^F(L=2,\lambda=2)$. 
The energies of the XXZ chain are separated according to the $S^z$ 
sector and those of the Ising chain are separated according to their 
parity $C=\pm 1$. The corresponding spin $S$ of the levels are shown 
and the energies forming $ V_0$ and $ V_1$ have a superscript 0 and 1, 
respectively. The energy level with an asterisk does not belong to 
$V_1$ since it degenerates with another energy with $S= 2$. From 
this table the exact correspondence between inhomogeneous $LM(3, \, 4)$ 
and the Ising chain is clear.

Thus the relations between the two models are very similar 
those for the critical homogeneous models.

\section*{Acknowledgments}

We are grateful to P. Martin, B. Feigin and especially to 
P. Pearce for usefull discussions and also to W. Everett 
for his help in the preparation of the text of this paper.
This work is supported by the following grants: 
RFBR-00-15-96579, RFBR-01-02-16686, INTAS-00-00055.

\section*{Appendices}

\appendix

\section{The limit of Sklyanin's transfer matrix}
\label{C}

The limit of the monodromy matrix $L(u)$ as 
$u\rightarrow -i \infty$ equals
$$
L(u \rightarrow -i \infty)={(-i)}^N2^{-N}e^{i(u+{\eta}/2)(N-1)}
\left (
\begin{array}{cc}
e^{i(u+{\eta}/2)}q^H & (q-q^{-1})X_0 \\
\vspace{-2mm}\\
(q-q^{-1})X & e^{i(u+{\eta}/2)}q^{-H}
\end{array}
\right ),
$$
where
\begin{equation}
\begin{array}{l}
H=\displaystyle \sum_{n=1}^{N}\frac{{\sigma}^z_n}{2},  
\;\;\;\; q=e^{i{\eta}},\\
X_0=\displaystyle \sum_{n=1}^{N}q^{-\frac12
({\sigma}^z_1+...+{\sigma}^z_{n-1})} {\sigma}^{-}_n
q^{-\frac12 ({\sigma}^z_{n+1}+...+{\sigma}^z_{N})}.
\end{array}
\end{equation}  
We can write our transfer matrix as:
\begin{equation}
T_{1/2}(u)={(-1)}^N\displaystyle \sum_{m,n=1}^{2}q^{2m-3}e^{2i(m-n)u}
L_{m,n}(u)L^t_{m,n}(u)
\label{eq:transfer_Sklyanin}
\end{equation}
Where the indices (m,n) are the indices of the elments of $L(u)$ 
and $L^t_{m,n}$ is an operator in "quantum" space 
${({\bf C}^2)}^N$ given by transposition of $L_{m,n}$. 
We find the limit $T(u \rightarrow -i \infty)$. Using 
(\ref{eq:transfer_Sklyanin}), we obtain
\begin{equation}
\begin{array}{lll}
T_{1/2}(u \rightarrow -i \infty) &=&
{(-1)}^N2^{-2N} q^{N-1} e^{2iu(N-1)} \\ 
\vspace{-2mm}\\
& &\times
\left( q^{-1}L_{1,1}L^t_{1,1}+q L_{2,2} L^t_{2,2}+q e^{2iu}L_{2,1}
L^t_{2,1}\right.\\
\vspace{-2mm}\\
&& \qquad + \left. q^{-1} e^{-2iu}L_{1,2}L^t_{1,2}\right)\\
\vspace{-2mm}\\
&=& 2^{-2N} q^{N-1} e^{2iu(N-1)}\\
\vspace{-2mm}\\
&  &\times \left(
q^{-1}e^{2iu}q q^{2H}+q e^{2 i u}(q-q^{-1})^2 X X^t
\right.\\
\vspace{-2mm}\\
&& \qquad + \left. q^{-1} e^{-2iu}{(q-q^{-1})}^2 X_0 X_0^t
+qe^{2iu} q q^{-2H}
\right) \end{array}
\end{equation}
The last term vanishes exponentially, and we can therefore 
ignore it. Using $X^t=Y$ (see the previous section), we obtain
\begin{equation}
T_{1/2}(u \rightarrow -i \infty) = 2^{-2N} q^{N} e^{2iNu} \left\{
q^{-1}q^{2H}+q q^{-2H}+ {(q-q^{-1})}^2 X Y \right\}.
\end{equation}
Using
$$
X Y = Y X + \frac{q^{2H}-q^{-2H}}{q-q^{-1}},
$$
we obtain
\begin{equation}
T_{1/2}(u \rightarrow -i \infty)=2^{-2N} q^{N} e^{2iNu} \left\{
{(q-q^{-1})}^2
{(S^2)}_q + (q+ q^{-1})\right\},
\end{equation}
where ${(S^2)}_q$ is Casimir of the algebra $U_q(sl(2))$:
$$
{(S^2)_q}=Y X + {\left( \frac{q^{H+1/2}-q^{-H-1/2}}{q-q^{-1}} \right)}^2-
{\left( \frac{q^{1/2}-q^{-1/2}}{q-q^{-1}} \right)}^2.
$$

\section{The operator $C$ and the limit of the transfer 
matrix of the Ising model with free boundary conditions.}
\label{D}

We prove that
$$
C=\frac{1}{2^{L+1}}T_{\rm Ising}(-i \infty),
$$
where $T_{\rm Ising}(u)$ is the transfer matrix of the Ising model.

In the case with free boundary conditions, an element of the 
transfer matrix equals
\begin{equation}
\begin{array}{l}
\displaystyle
\sum_{\sigma''}\exp\left[J{\sigma}{''_1}({\sigma}_1+{\sigma}{'_1})\right]
\prod_{j=1}^{L}
\exp\left[K{\sigma}{''_j}({\sigma}_j+{\sigma}{'_j})+
J{\sigma}{''_{j+1}}({\sigma}_j+{\sigma}{'_j})\right]\\
\qquad \qquad \qquad \times \displaystyle
\exp\left[K{\sigma}{''_{L+1}}({\sigma}_{L}+{\sigma}{'_{L}})\right]\\
= \displaystyle 2^{L+1}
\prod_{j=1}^{L}\cosh\left[J({\sigma}_1+{\sigma}{'_1})\nonumber\right]
\cosh\left[K({\sigma}_j+{\sigma}{'_j})+J({\sigma}_j+{\sigma}{'_j})\right]
\\
\qquad \qquad \phantom{{}={}} \displaystyle \times
\cosh\left[K({\sigma}_{L}+{\sigma}{'_{L}})\right]
\end{array}
\end{equation} 
For $u=-i \infty$,
$$
\cosh(2J)=0, \qquad \cosh(2K)=0.
$$
Hence, if there is even one pair ${\sigma}_j$, ${\sigma}{'_j}$ 
such that ${\sigma}_j={\sigma}{'_j}$, then 
the matrix element equals zero. We can see from this 
that $T(-i \infty)$ is actually proportional to the 
product of all ${\sigma}^x_i$ (with the coefficient of 
proportionality equaling $2^{L+1}$), as this product has 
only those elements nonzero for which ${\sigma}_i=-
{\sigma}{'_i}$ for all $i$. We obtain hence the desired 
result.

\newpage
\Large
\begin{center}
Table Captions
\end{center}
\normalsize
\vspace{2.0cm}
 
Table 1 - Eigenenergies of the normalized Hamiltonians
$\tilde H_{XXZ}$ and $\tilde H_{\rm Ising}^M$ given by 
(\ref{eq:Ising_tilde}) and (\ref{eq:XXZ_tilde}). 
The  eigenenergies of the XXZ Hamiltonian are separated 
into the sectors labelled by the z-magnetization $S^z$. 
The spins $S$ of the highest weights are also shown. 
The levels marked by $(+)$ form $LM(3, \, 4)$ and coincide 
with those of the Ising quantum chain. The level marked by $(*)$ 
does not belong to $LM(3, \, 4)$ since it is degenerated 
with another level with $S =5/2$.
\vspace{1cm}

Table 2 - Part of the eigenenergies of the normalized 
Hamiltonians $\tilde H_{XXZ}$ given by (\ref{eq:XXZ_tilde}) 
and the corresponding ones of the $\tilde H_{\rm Ising}^{++}$ 
and $\tilde H_{\rm Ising}^{+-}$ given by (\ref{eq:Ising_tilde}).
The eigenenergies of the XXZ Hamiltonian are separated into 
the sectors labelled by the z-magnetization $S^z$. The spins 
$S$ of the highest weights are also shown. The levels with the 
superscript $(0)$ and $(1)$ belongs to the sectors $V_0$ and 
$V_1$ of $LM(3, \, 4)$, respectively. The levels marked by $(*)$ 
does not belong to $LM(3, \, 4)$ since they are degenerated 
with others levels with $S=2$.
\vspace{1cm}

Table 3 - Eigenenergies of the normalized Hamiltonians
$\tilde H_{XXZ}$ and $\tilde H_{\rm Ising}^{F}$ given by 
(\ref{eq:Ising_tilde}) and (\ref{eq:XXZ_tilde}), 
respectively. The energies of the XXZ Hamiltonian are 
separated into the sectors labelled by the z-magnetization 
$S^z$, and those of the Ising chain are separated 
according to their parity $C = \pm 1$. The spins $S$ of the 
highest weights of the XXZ Hamiltonian are also shown. The 
levels with the superscript $(0)$ and $(1)$ belongs to the 
sectors $V_0$ and $V_1$ of $LM(3, \, 4)$, respectively. 
The level marked by $(*)$ does not belong to $LM(3, \, 4)$ 
since it is degenerated with another level with $S=2$.
\newpage
\vspace{2cm}
\Large
Table 1
\normalsize
\vspace{1cm}
\begin{center}

\begin{tabular}{|c|c|c|c|c|}    \hline
&        \multicolumn{3} {c|} {$\tilde H_{XXZ}(N=5,\lambda=2)$} &
$\tilde H_{\rm Ising}^M(L=2,\lambda=2)$  \\ \hline
$S$ & $S^z =\pm 1/2$ & $S^z = \pm 3/2$ &
$S^z = \pm 5/2$   & \\ \hline
$1/2$ & $-7.33821^{(+)}$ &             &  &   -7.33821 \\
$1/2$ & $-4.88872^{(+)}$ &             &  &   -4.88872 \\
$3/2$ & -4.51633       & -4.51633    &  &            \\
$1/2$ & $-3.59656^{(+)}$ &             &  &   -3.59656 \\
$3/2$ & -3.24557       & -3.24557    &  &            \\
$1/2$ & $-1.14707^{(+)}$ &             &  &   -1.14707 \\
$3/2$ & -0.99797       & -0.99797    &  &            \\
$5/2$ & 0              & 0           & 0&            \\
$1/2$ & 0 (*)          &             &  &            \\
$3/2$ & 0.27369        & 0.27369     &  &            \\ \hline
\end{tabular}
\end{center}
\Large
\vspace{1cm}
Table 2
\normalsize
\begin{center}
 
\begin{tabular}{|c|c|c|c|c|c|}  \hline
&        \multicolumn{3} {c|} {$\tilde H_{XXZ}(N=6,\lambda=2)$} &
$\tilde H_{\rm Ising}^{++}(L=2,\lambda=2)$ &
$\tilde H_{sing}^{+-}(L=2,\lambda=2)$  \\ \hline
$S$ & $S^z =0$ & $S^z = \pm 1$ &
$S^z = \pm 2$ &  &   \\ \hline
0 & $-8.34840^{(0)}$ &             &  &   -8.34840 &  \\
1 & $-8.07413^{(1)}$ & $-8.07413^{(1)}$ &     &  &   -8.07413   \\
1 & $-5.65685^{(1)}$ & $-5.65685^{(1)}$ &     &  &   -5.65685   \\
0 & $-5.38259^{(0)}$ &             &  &   -5.38259 &  \\
2 & -4.67083 & -4.67083 & -4.67083    &  &      \\
1 & -4.67083(*) & -4.67083(*) &     &  &      \\
1 & $-4.51691^{(1)}$ & $-4.51691^{(1)}$ &     &  & -4.51691      \\
0 & $-4.24264^{(0)}$ &             &  &   -4.24264 &  \\
2 & -3.70246 & -3.70246 & -3.70246    &  &      \\
1 & -3.70246(*) & -3.70246(*) &     &  &      \\
0 & $-1.82356^{(0)}$ &             &  &   -1.82356 &  \\
1 & $-1.55109^{(1)}$ & $-1.55109^{(1)}$ &     &  & -1.55109      \\ \hline
\end{tabular}
\end{center}
\Large
\newpage
Table 3
\normalsize
\vspace{1cm} 
\begin{center}

\begin{tabular}{|c|c|c|c|c|c|}  \hline
&        \multicolumn{3} {c|} {$\tilde H_{XXZ}(N=4,\lambda=2)$} &
\multicolumn{2} {c|} {$\tilde H_{\rm Ising}^{F}(L=2,\lambda=2)$}
\\ \hline
$S$ & $S^z =0$ & $S^z = \pm 1$ &
$S^z = \pm 2$ & $C = +1$ & $C = -1$   \\ \hline
0 & $-6.4510^{(0)}$ &             &  & -6.45101  &  \\
1 & $-4.24264^{(1)}$ & $-4.24264^{(1)}$ &     &  &   -4.24264   \\
1 & $-2.82843^{(1)}$ & $-2.82843^{(1)}$ &     &  &   -2.82843   \\
0 & $-0.62006^{(0)}$ &             &  &   -0.62006 &  \\
1 & 0(*) & 0(*) &     &  &      \\
2 & 0 & 0 & 0    &  &      \\ \hline
\end{tabular}
\end{center}

\begin{thebibliography}{3}
\bibitem{kn:gnus1}
 Behrend R E,  Pearce P A and  O'Brien D L  
{ Interaction-round-a-face models with fixed boundary conditions: The ABF
fusion hierarchy}, {\it J. Stat. Phys.} {\bf 84} 1  (1996)
\bibitem{kn:gnus2}
 O'Brien D L ,  Pearce P A  and  Warnaar S O  { Finitized conformal
spectrum of the Ising model on the cylinder and torus.} {\it Physica A} \
{\bf 228} 63-77 (1996)
\bibitem{kn:gnus3}
 Alcaraz F C , Barber M N and   Batchelor M T  
Conformal invariance, the XXZ chain and the 
operator content of two-dimensional 
critical systems
{ it Ann. Phys. (N.Y.)} {\bf 182} 280-343 (1988) ;
 Alcaraz F C,   Barber M N,  Batchelor M T,  Baxter R J and  
 Quispel G R W  Surface exponents of the quantum XXZ, Ashkin-Teller 
and Potts models {\it J. Phys. A: Math. Gen. } {\bf 20}
 6397-6409 (1987)
\bibitem{kn:gnus4}
 Pasquier V and   Saleur H Common structures between finite systems and 
conformal field-theories through quantum groups {\it Nucl. Phys. B }
{\bf 330} 523 (1990)
\bibitem{kn:gnus6}
 Belavin A and Yu. Stroganov Yu
{ Minimal Models of Integrable Lattice Theory and Truncated Functional
Equations} {\it Phys. Lett. B} {\bf 446} 281 (1999)
\bibitem{kn:gnus7}
Lusztig E {\it Contemp. Math.} {\bf 82}  59 (1989)
\bibitem{kn:gnus9}
 Hamer C J Q-state Potts models in Hamiltonian field theory for 
Q-greater-than-or-equal-to-4 in (1+1) dimensions  
{\it J. Phys. A: Math. Gen} {\bf 14}  2981 (1981)
\bibitem{kn:gnus11}
 Sklyanin E K  { Boundary conditions for integrable quantum systems}, 
{\it J. Phys. A: Math. Gen.} {\bf 21}  2375-2389 (1988)
\bibitem{kn:gnus12}
 Kulish P P  and Sklyanin E K { The general $U_q[sl(2)]$ invariant XXZ
integrable quantum spin chain} {\it J. Phys. A: Math. Gen.} {\bf 24} 
L435-L439 (1991)
\bibitem{kn:gnus13}
 Martin P P, { Potts models and related problems in statistical
mechanics} (World Scientific, Singapore) 1991.
\bibitem{Alc}
 Alcaraz F C,  Baake M,  Grimm U and  Rittenberg V,
{ The modified XXZ Heisenberg chain, conformal 
invariance and the surface exponents of $c<1$ systems}, 
{\it J. Phys. A: Math. Gen.} {\bf 22} L5-L11 (1989) 
\bibitem{kn:gnus14}
 Belavin A A and  Usmanov R A {\it to be published}
\end{thebibliography}
\end{document}